\documentclass[twocolumn,superscriptaddress,preprintnumbers,tightenlines,nofootinbib,showpacs,eqsecnum,amsfonts,amsmath]{revtex4}
\usepackage{enumerate}
\usepackage{graphicx}
\usepackage[caption=false, justification=centerlast]{subfig}
\usepackage{enumitem,amssymb}
\usepackage{cprotect}
\usepackage{xcolor}
\usepackage{fancyvrb}
\usepackage{hyperref}
\usepackage{siunitx}
\usepackage{url}
\usepackage{bm}

\pagecolor{white}

\newcommand{\dd}{\mathrm{d}}
\newcommand{\phenom}{\texttt{\detokenize{IMRPhenom}}}
\newcommand{\teobr}{\texttt{\detokenize{TEOBResumS}}}
\newcommand{\seobnr}{\texttt{\detokenize{SEOBNR}}}
\newcommand{\seobnrvfour}{\texttt{\detokenize{SEOBNRv4}}}
\newcommand{\hmpa}{\texttt{\detokenize{SEOBNRv4HM_PA}}}
\newcommand{\hmm}{\texttt{\detokenize{SEOBNRv4HM}}}
\newcommand{\hmmp}{\texttt{\detokenize{SEOBNRv4PHM}}}
\newcommand{\hmrom}{\texttt{\detokenize{SEOBNRv4HM_ROM}}}
\newcommand{\hmopt}{\texttt{\detokenize{SEOBNRv4_opt}}}
\newcommand{\sur}{\texttt{\detokenize{SEOBNRv4T_surrogate}}}
\newcommand{\event}{\texttt{\detokenize{GW190412}}}
\newcommand{\tgrevent}{\texttt{\detokenize{GW150914}}}
\newcommand{\lalsuite}{\texttt{\detokenize{LALSuite}}}
\newcommand{\lalsim}{\texttt{\detokenize{LALSimulation}}}
\newcommand{\lalmcmc}{\texttt{\detokenize{LALinferenceMCMC}}}
\newcommand{\gw}{\textsc{gw}}
\newcommand{\nr}{\textsc{nr}}
\newcommand{\ligo}{\textsc{ligo}}
\newcommand{\lvk}{\textsc{lvk}}

\newcommand{\lisa}{\textsc{lisa}}
\newcommand{\lal}{\textsc{lal}}
\newcommand{\eob}{\textsc{eob}}
\newcommand{\gr}{\textsc{gr}}
\newcommand{\pa}{\textsc{pa}}
\newcommand{\ode}{\textsc{ode}}
\newcommand{\pe}{\textsc{pe}}
\newcommand{\bh}{\textsc{bh}}
\newcommand{\bbh}{\textsc{bbh}}
\newcommand{\bns}{\textsc{bns}}
\newcommand{\qnm}{\textsc{qnm}}
\newcommand{\psd}{\textsc{psd}}
\newcommand{\snr}{\textsc{snr}}
\newcommand{\js}{\textsc{js}}
\newcommand{\bigo}[0]{\mathcal{O}}
\newcommand{\chieff}[0]{\chi_{\mathrm{eff}}}

\newcommand{\pSEOB}{\texttt{pSEOBNR}}

\def\qSEOBNRv4HM{$0.39_{-0.06}^{+0.08}$}
\def\qSEOBNRv4HMmedian{$0.39$}
\def\qSEOBNRv4HMupper{$0.47$}
\def\qSEOBNRv4HMlower{$0.33$}
\def\qSEOBNRv4HM_PA{$0.38_{-0.05}^{+0.08}$}
\def\qSEOBNRv4HM_PAmedian{$0.38$}
\def\qSEOBNRv4HM_PAupper{$0.46$}
\def\qSEOBNRv4HM_PAlower{$0.33$}
\def\d_LSEOBNRv4HM{$1652.37_{-429.35}^{+468.47}$}
\def\d_LSEOBNRv4HMmedian{$1652.37$}
\def\d_LSEOBNRv4HMupper{$2120.84$}
\def\d_LSEOBNRv4HMlower{$1223.02$}
\def\d_LSEOBNRv4HM_PA{$1671.65_{-445.83}^{+458.36}$}
\def\d_LSEOBNRv4HM_PAmedian{$1671.65$}
\def\d_LSEOBNRv4HM_PAupper{$2130.01$}
\def\d_LSEOBNRv4HM_PAlower{$1225.82$}
\def\thetajnSEOBNRv4HM{$0.88_{-0.34}^{+0.29}$}
\def\thetajnSEOBNRv4HMmedian{$0.88$}
\def\thetajnSEOBNRv4HMupper{$1.17$}
\def\thetajnSEOBNRv4HMlower{$0.54$}
\def\thetajnSEOBNRv4HM_PA{$0.87_{-0.34}^{+0.3}$}
\def\thetajnSEOBNRv4HM_PAmedian{$0.87$}
\def\thetajnSEOBNRv4HM_PAupper{$1.17$}
\def\thetajnSEOBNRv4HM_PAlower{$0.53$}

\def\qSEOBNRv4HM_PA{$0.28_{-0.05}^{+0.07}$}
\def\qSEOBNRv4HM_PAmedian{$0.28$}
\def\qSEOBNRv4HM_PAupper{$0.35$}
\def\qSEOBNRv4HM_PAlower{$0.23$}
\def\qSEOBNRv4HM_ROM{$0.28_{-0.05}^{+0.07}$}
\def\qSEOBNRv4HM_ROMmedian{$0.28$}
\def\qSEOBNRv4HM_ROMupper{$0.35$}
\def\qSEOBNRv4HM_ROMlower{$0.23$}
\def\d_LSEOBNRv4HM_PA{$698.39_{-190.57}^{+182.25}$}
\def\d_LSEOBNRv4HM_PAmedian{$698.39$}
\def\d_LSEOBNRv4HM_PAupper{$880.64$}
\def\d_LSEOBNRv4HM_PAlower{$507.82$}
\def\d_LSEOBNRv4HM_ROM{$697.58_{-190.63}^{+181.28}$}
\def\d_LSEOBNRv4HM_ROMmedian{$697.58$}
\def\d_LSEOBNRv4HM_ROMupper{$878.86$}
\def\d_LSEOBNRv4HM_ROMlower{$506.95$}
\def\thetajnSEOBNRv4HM_PA{$0.84_{-0.3}^{+0.47}$}
\def\thetajnSEOBNRv4HM_PAmedian{$0.84$}
\def\thetajnSEOBNRv4HM_PAupper{$1.31$}
\def\thetajnSEOBNRv4HM_PAlower{$0.54$}
\def\thetajnSEOBNRv4HM_ROM{$0.83_{-0.29}^{+0.39}$}
\def\thetajnSEOBNRv4HM_ROMmedian{$0.83$}
\def\thetajnSEOBNRv4HM_ROMupper{$1.22$}
\def\thetajnSEOBNRv4HM_ROMlower{$0.54$}

\newcommand{\sig}{\mathrm{\textsc{s}}}
\newcommand{\temp}{\mathrm{\textsc{t}}}

\newcommand{\overbar}[1]{\mkern 1.5mu\overline{\mkern-1.5mu#1\mkern-1.5mu}\mkern 1.5mu}

\def\domega220JS150914{$1.02\times 10^{-3}$}
\def\dtau220JS150914{$4.4\times 10^{-4}$}

\newcommand{\AEI}{\affiliation{Max Planck Institute for Gravitational Physics (Albert Einstein Institute), Am M\"{u}hlenberg 1, Potsdam 14476, Germany}}
\newcommand{\Maryland}{\affiliation{Department of Physics, University of Maryland, College Park, MD 20742, USA}}

\hypersetup{
	colorlinks=true,
    linkcolor=cyan,
    citecolor=cyan,
    filecolor=magenta,      
    urlcolor=cyan,
}

\allowdisplaybreaks

\graphicspath{{./Figures/}}

\begin{document}

\pagenumbering{arabic}

\title{Fast post-adiabatic waveforms in the time domain:\\Applications to compact binary coalescences in LIGO and Virgo}

\author{Deyan P. Mihaylov}
\AEI
\email{deyan@aei.mpg.de}

\author{Serguei Ossokine}
\AEI

\author{Alessandra Buonanno}
\AEI
\Maryland

\author{Abhirup Ghosh}
\AEI

\date{\today}

\begin{abstract}
  We present a computationally efficient (time-domain) multipolar
  waveform model for quasi-circular spin-aligned compact binary coalescences. The
  model combines the advantages of the numerical-relativity informed,
  effective-one-body (\textsc{eob}) family of models with a post-adiabatic solution
  of the equations of motion for the inspiral part of the two-body
  dynamics. We benchmark this model against other state-of-the-art
  waveforms in terms of efficiency and accuracy. We find a speed-up of
  one to two orders of magnitude compared to the underlying
  time-domain \textsc{eob} model for the total mass range \(2 - 100
  M_{\odot}\). More specifically, for a low total-mass system, such as a
binary neutron star with equal masses of \(1.4 M_{\odot}\), like \texttt{GW170817}, the computational speedup is
  around 100 times; for an event with total mass \(\sim 40 M_\odot\) and mass 
ratio \(\sim 3\), like \texttt{GW190412}, the speedup is
  by a factor of \(\sim 20\), while for a binary system of comparable masses and total mass 
of \(\sim 70 M_{\odot}\), like \texttt{GW150914}, it is by a factor of \(\sim 10\). 
We demonstrate that the new model is extremely faithful
  to the underlying \textsc{eob} model with unfaithfulness less than
  \(0.01\%\) across the entire applicable region of parameter
  space. Finally, we present successful applications of this new
  waveform model to parameter estimation studies and tests of general
  relativity.
\end{abstract}


\maketitle

\section{Introduction}
\label{sec:Intro}
\noindent

Since 2015, the detections of gravitational waves (\gw s) have yielded a
wealth of remarkable discoveries \citep{Abbott:2016blz,
  TheLIGOScientific:2016pea, Venumadhav:2019lyq, Nitz:2018imz,
  Nitz:2019hdf, Zackay:2019btq, LIGOScientific:2018mvr, Abbott:2020niy}. In the three
observing runs \citep{LIGOScientific:2018mvr, Abbott:2020niy} of the
Advanced \ligo\ \citep{TheLIGOScientific:2014jea} and Advanced Virgo
\citep{TheVirgo:2014hva} detectors, a total of 50 events have been
observed and confirmed; among these are both binary black hole (\bbh)
mergers and binary neutron star (\bns) mergers
\citep{detections}. Particularly interesting are the discoveries of
binaries \texttt{\detokenize{GW190412}} with mass ratio 3
\citep{LIGOScientific:2020stg} and \texttt{\detokenize{GW190814}}
(which could be the first ever detected merger of a black hole and a
neutron star) \citep{Abbott:2020khf} with mass ratio
10. \texttt{\detokenize{GW190521}} is the most massive binary detected
so far with a total mass of \(150 M_{\odot}\) \citep{Abbott:2020tfl}. 

Detections of compact binary mergers are expected to increase in the
coming years \citep{Dominik:2014yma, Belczynski:2015tba}: during the
upcoming \ligo\ and Virgo observing runs~\citep{Abbott:2020qfu}, and
with future ground-based detectors like the Einstein Telescope
\citep{Punturo:2010zz} and Cosmic Explorer
\citep{Reitze:2019iox}, and the space-based mission \lisa~\citep{Audley:2017drz}. Extracting information from such \gw\
detections relies on accurate and computationally efficient models of
the gravitational waveforms, which are emitted during coalescence
\citep{Buonanno:2014aza,LIGOScientific:2019hgc}. Firstly, the estimation of the binary
parameters of a typical event (using Bayesian inference, Markov
chains, or similar methods) requires on the order of several million
evaluations of the waveform models \citep{Harry:2009ea, Veitch:2014wba}. On the other hand, the \gw\ phase needs to be
accurate to less than a cycle of the binary in order to avoid
ambiguity in the estimations \citep{Cutler:1992tc, Lindblom:2008cm}. Upcoming runs, as well as future
detectors, will require even better waveform accuracy in order to
reliably identify and analyse \gw\ events~\citep{Purrer:2019jcp}. Accurately identifying the properties
of a large population of binaries will allow us to make inferences
on scenarios of compact-object binary formation~\citep{Abbott:2020gyp}, and also carry out more stringent tests 
of General Relativity (\gr) in the highly dynamical, strong-field regime~\citep{Abbott:2020jks}. For these reasons,
work on more advanced and innovative waveform models continues for
\ligo, Virgo, and future \gw\ missions.

Gravitational-wave models that include the inspiral, merger, and ringdown stages of a
compact binary coalescence have been developed using the effective-one-body (\eob) formalism~\citep{Buonanno:1998gg, Buonanno:2000ef, 
Damour:2000we, Damour:2001tu, Buonanno:2005xu, Buonanno:2006ui, Damour:2008gu} 
(notably the \seobnr~\citep{Pan:2010hz, Pan:2011gk, Barausse:2009xi, Taracchini:2012ig, Taracchini:2013rva,
  Pan:2013rra,Boh__2017,Babak:2016tgq, Cotesta:2018fcv, Ossokine:2020kjp} 
 and \teobr~\citep{Damour:2014sva, Nagar:2015xqa,Nagar:2018zoe,Nagar:2021gss} waveform models), 
and the inspiral-merger-ringdown phenomenological approach~\citep{Ajith:2009bn, Santamaria:2010yb, Hannam:2013oca, Husa:2015iqa,
  Khan:2015jqa, London:2017bcn, Khan:2018fmp, Khan:2019kot, Pratten:2020fqn} (i.e., the \phenom\ models). The \eob\ families of models employ a
semi-analytic approach that combines an analytical description of the spinning two-body 
dynamics and gravitational radiation for the entire 
coalescence with numerical relativity (\nr) information in the strong-field
merger-ringdown regime~\citep{Pretorius:2005gq, Campanelli:2005dd, Baker:2005vv, Brugmann:2008zz, Centrella:2010mx,
  Mroue:2013xna, Jani:2016wkt, Healy:2017psd, Boyle:2019kee}. Here, we will focus on the SEOBNR waveform models, which 
were used, so far, by LIGO and Virgo detectors to observe GW signals and infer astrophysical properties 
(e.g., see Refs.~\citep{LIGOScientific:2018mvr, Abbott:2020niy}) and test GR (e.g., see Ref.~\citep{Abbott:2020jks}).

The time-domain \seobnr\ models~\citep{Boh__2017,Cotesta:2018fcv,Ossokine:2020kjp} are routinely employed in data analysis 
for sufficiently high-mass binaries with the \ligo\ Algorithm Library (\lal) Inference codes~\citep{Veitch:2014wba}, while, for generic-mass binaries, fast parameter-estimation codes are required~\citep{Lange:2018pyp}. Nevertheless, the time for generating a waveform in the
low-mass regime (\(\lesssim 5 M_{\odot}\)) can be on the order of \(\sim \SI{100}{\second}\) or even longer starting at \SI{20}{\hertz}. 
Thus, there are parts of the binary's parameter space, for which the time-domain \seobnr\ models are not suitable 
for direct use in parameter estimations without further optimizations, like the ones
discussed in the current publication, which was originally introduced in Refs.~\citep{Nagar:2018gnk, Rettegno:2019tzh}.
Alternative methods have been developed in order to afford speedy data
analysis for \gw s. Reduced order modeling and surrogate techniques
\citep{Field:2013cfa, Purrer:2014fza, Blackman:2015pia,
  Purrer:2015tud, Blackman:2017dfb, Doctor:2017csx, Blackman:2017pcm,
  Lackey:2018zvw, Setyawati:2019xzw, Cotesta:2020qhw} have been
successfully applied to \eob\ waveform models \citep{Field:2013cfa,
  Purrer:2014fza, Purrer:2015tud, Boh__2017, Lackey:2018zvw} and to
pure \nr-waveforms \citep{Blackman:2015pia, Blackman:2017dfb,
  Blackman:2017pcm, Varma:2018mmi, Varma:2019csw}. These methods work by
decomposing and interpolating the waveforms on a sparse grid in time
or frequency domain, and then using interpolation or more
sophisticated statistical methods to obtain the fitting parameters
across the binary's parameter space under study. The resulting waveform model
is then verified for accuracy against an independent testing
set. However, although very successful, such models suffer from certain limitations. By construction, 
they are restricted to confined regions of the parameter space, and
have to be developed from scratch if the underlying time-domain model
is updated --- for example when more physical effects are included or 
higher-order post-Newtonian parameters are added to make 
these waveforms more accurate.

Here, we develop the multipolar \hmpa\ waveform model for
spin-aligned compact binaries moving on quasi-circular orbits. This model is a computationally cheaper version
of the time-domain \hmm\ waveform model \citep{Cotesta:2018fcv} and, as such, it includes
higher-order harmonics (or higher modes, \textsc{hm}), which are important for asymmetric mass-ratio
binaries, high-mass and high-inclination systems \citep{Brown:2012nn,
  Capano:2013raa, Varma:2014jxa, Varma:2016dnf, Harry:2017weg,
  Kalaghatgi:2019log}. In the \hmpa\ waveform model, the binary
dynamics is solved using a post-adiabatic (\pa) approach. The latter was 
proposed and applied to the \texttt{TEOBResumS} model in Refs.~\citep{Damour:2012ky,Nagar:2018gnk, Rettegno:2019tzh} 
and used in all subsequent publications (see, e.g., Ref.~\citep{Riemenschneider:2021ppj} and references therein). It was also 
implemented in \lalsim. In the \pa\ method, the inspiral evolution 
(until the last few orbits before merger) is approximated by an adiabatic solution of the (ordinary differential) equations of motion, 
with post-adiabatic corrections added iteratively up to the order needed 
to achieve the desired accuracy. In this work we apply this technique to construct
a fast and accurate aligned-spin dynamics  based on the 
\seobnrvfour\ model \citep{Boh__2017} and implement it in \lalsim. The 
speed-up and accuracy benchmarks are supported by applications of the \pa\ waveform
model for parameter estimation studies and tests of \gr.

The paper is organized as follows. Section~\ref{sec:theory} of this article reviews the \eob\ 
dynamics in the \pa\ approximation for arbitrary
Hamiltonians. Section~\ref{sec:implementation} presents the
implementation of this method in the \lalsim\ waveform model library
(as approximant \hmpa), and benchmarks the model against other
established \eob\ models. Section~\ref{sec:pestudy} presents two
parameter estimation (\textsc{pe}) studies using the \hmpa\ model. In
Section~\ref{sec:tgr} the model is applied to a ringdown test of
\gr. Section~\ref{sec:conclusion} concludes the article with a
discussion of the significance of this work and its possible future
directions. We shall work in natural units \(G=1=c\).

\section{Post-adiabatic approximation to the inspiral dynamics}
\label{sec:theory}
\noindent

The \eob\ formalism provides an analytical description of the \gw\
emission from the process of binary coalescence, including inspiral,
merger, and ringdown~\citep{Buonanno:1998gg, Buonanno:2000ef}. The accuracy of this description can be further
improved by calibrating against \nr\ simulations. 

A binary system composed of two \bh s moving on a quasi-circular orbit 
with spins aligned or anti-aligned (henceforth, spin-aligned for short) 
with the orbital angular momentum is described by four parameters: the component masses \(m_1\) and \(m_2\), and the 
(dimensionless) spins \(\chi_1 = S_1/m_1^{2}\) and \(\chi_2 = S_2/m_{2}^{2}\). In the \eob\
approach the (center-of-mass) two-body dynamics is mapped onto the dynamics of 
an effective body of mass \(\mu = m_{1} m_{2} / (m_{1} + m_{2})\), which 
moves in a deformed Kerr spacetime of mass \(M =
m_{1} + m_{2}\), the deformation parameter being the symmetric mass 
ratio \(\nu = \mu/M\). The conservative two-body dynamics is obtained from the \eob\
Hamiltonian~\citep{Buonanno:1998gg, Barausse:2009xi}:
\begin{align}
H = M \sqrt{1 + 2 \nu \left(\!\frac{H_{\mathrm{eff}}}{\mu}-1\!\right)} - M,
\end{align}
where \(H_{\mathrm{eff}}\) is the Hamiltonian that describe the motion of the effective body of mass \(\mu\) and spin \(S_{*} = [(m_{2}/m_{1}) S_{1} + (m_{1}/m_{2}) S_{2}] / M^{2}\) in the (deformed) Kerr spacetime of mass \(M\) with spin \(S = S_{1} + S_{2}\).

For aligned-spin binaries, the motion is constrained to a fixed plane. Thus, we use polar coordinates and introduce the phase-space 
dimensionless variables \((r, \varphi, p_{r_{*}}, p_{\varphi})\) related to the physical ones through the following expressions
\begin{subequations}
\begin{align}
r = \frac{R}{M}, \quad p_{r_{*}} = \frac{P_{R_{*}}}{\mu}, \quad p_{\varphi} = \frac{P_{\varphi}}{\mu M}.
\end{align}
\end{subequations}
The radial momentum \(p_{r_{*}}\) is conjugate to the tortoise coordinate of the deformed spacetime \(r_{*}\) \citep{Damour:2007yf,Pan:2011gk}. The dissipative effects in the \eob\ formalism 
are described by the radiation-reaction force~\citep{Buonanno:2000ef, Buonanno:2006ui, Damour:2008gu, Pan:2010hz}
\begin{align}
\bm{\mathcal{F}} = \frac{\Omega}{16 \pi} \frac{\bm{p}}{|\bm{L}|} \sum_{\ell = 2} \sum_{m = -\ell}^{\ell} m^{2} \left|d_{L}\,h_{\ell m}\right|^{2},
\end{align}
where \(\Omega\) is the angular orbital frequency, \(\bm{L}\) is the orbital angular momentum, \(d_{L}\) is the luminosity distance and \(h_{\ell m}\) are the gravitational 
modes far from the source. In this setup, the equations of motion read \citep{Pan:2011gk}:
\begin{subequations}\label{eq:eobeom}
\begin{align}
\frac{\dd r}{\dd t} &= \frac{\dd p_{r_{*}}}{\dd p_{r}} \, \frac{\partial H}{\partial p_{r_{*}}}, \label{eq:r_eom} \\
\frac{\dd \varphi}{\dd t} &= \frac{\partial H}{\partial p_{\varphi}}, \label{eq:phi_eom} \\
\frac{\dd p_{r_{*}}}{\dd t} &= - \frac{\dd p_{r_{*}}}{\dd p_{r}} \, \frac{\partial H}{\partial r} + \mathcal{F}_{r}, \label{eq:p_r_eom} \\
\frac{\dd p_{\varphi}}{\dd t} &= \mathcal{F}_{\varphi}. \label{eq:p_phi_eom}
\end{align}
\end{subequations}
Here, \(t = T / M\) is a dimensionless time variable.

The usual procedure employed in \eob\ waveform models involves solving
the eqs.~(\ref{eq:eobeom}) numerically, using an ordinary differential equation (\ode) integrator with
a suitable time step and initial conditions. This is often
computationally expensive (especially for longer waveforms), and is
one of the bottlenecks for efficiently generating the \eob\
waveform. The post-adiabatic (\pa) approximation~\citep{Buonanno:2000ef, Damour:2012ky, Nagar:2018gnk} converts the \ode\ equations of
motion into a set of non-linear algebraic equations which need to be
solved numerically, but have a lower computational cost associated
with them.

The adiabatic approximation assumes that the dynamics is comprised of
a sequence of circular orbits. As such, there is no radiation
reaction, hence \(\mathcal{F}_{\varphi} = 0\) and \(p_{r_{*}}\)
vanishes. Hence, the leading-order orbital angular momentum
\(p_{\varphi}\) can be calculated at a given radius from
Eq.~(\ref{eq:p_r_eom}):
\begin{align}
\left.\frac{\partial H}{\partial r}\right|_{p_{r_{*}} =\,0, p_{\varphi}, r} = 0. \label{eq:j0}
\end{align}

The post-adiabatic approximation assumes that the radiation reaction \(\mathcal{F}_{\varphi}\) which can be used to furnish \(p_{r_{*}}\) through a combination of Eqs.~(\ref{eq:r_eom}) and (\ref{eq:p_r_eom}):
\begin{align}
\frac{\dd p_{\varphi}}{\dd r} \, \frac{\partial H}{\partial p_{r_{*}}} - \mathcal{F}_{\varphi} &= 0. \label{eq:pa_pr}
\end{align}
At the post-post-adiabatic level, one can use the newly obtained approximation for \(p_{r_{*}}\) to additionally correct the orbital angular momentum \(p_{\varphi}\), this time utilising Eqs.~(\ref{eq:phi_eom}) and (\ref{eq:p_phi_eom}):
\begin{align}
\frac{\partial H}{\partial p_{r}} + \frac{\partial H}{\partial r} \, \frac{\dd r}{\dd p_{r_{*}}} - \frac{p_{r_{*}}}{p_{\varphi}} \, \mathcal{F}_{\varphi} &= 0. \label{eq:pa_pphi}
\end{align}

This approximation procedure can be iterated further, and the procedure for obtaining the corrections to the leading-order solution \((p_{r_{*}}, p_{\varphi}) = (0, j_{0}(r))\) can be formalised in the following way, as described in \citep{Nagar:2018gnk}. For each value of the radial coordinate \(r\), the radiation reaction can be written as an expansion in a formal parameter \(\epsilon\)
\begin{align}
\mathcal{F}_{\varphi} (r) = \sum_{n = 0}^{\infty} \epsilon^{2n+1} \mathcal{F}_{2n+1} (r)
\end{align}
Therefore, the solutions of the \eob\ equations of motion can also be written as an expansion in powers of this fictitious parameter:
\begin{subequations}
\begin{align}
p_{\varphi} (r) &= j_{0} (r) \left [1 + \sum_{n=1}^{\infty} \epsilon^{2n} \, \phi_{2n} (r) \right ]^{\!1/2}, \\
p_{r_{*}} (r) &= \sum_{n = 0}^{\infty} \epsilon^{2n+1} \, \rho_{2n+1} (r).
\end{align}
\end{subequations}
The \pa\ procedure allows for the two momenta to be calculated with arbitrary precision by adding more terms in the expansions above. The corrections at \((n, n+1)\)th \pa\ order can be found by iteratively solving Eq.~(\ref{eq:pa_pr}) for \(p_{r_{*}}\) and Eq.~(\ref{eq:pa_pphi}) for \(p_{\varphi}\). In solving these two equations, one must remember that all other variables (apart from the unknown one) must be kept at their most recent \pa\ order. This procedure can be repeated as many times as necessary, until the desired accuracy in terms of powers of \(\epsilon\) is achieved \citep{Nagar:2018gnk}.

In practice, we proceed as follows. As a start, a radial grid is constructed for the part of the two-body dynamics where the \pa\ approximation is to be applied, between two radii \(r_{\mathrm{max}}\) and \(r_{\mathrm{min}}\). At each node in this grid, the adiabatic solution \(j_{0} (r)\) is obtained through Eq.~(\ref{eq:j0}) --- it serves as the leading-order uncorrected values for the orbital angular momentum (the uncorrected value for \(p_{r_{*}}\) is chosen as 0 everywhere on the grid). To obtain the \(N\)-th order \pa\ approximation, the momenta \(p_{r_{*}} (r)\) and \(p_{\varphi} (r)\) are computed through Eqs.~(\ref{eq:pa_pr}) and (\ref{eq:pa_pphi}), respectively, at each point in the grid and this part is repeated up to the chosen \pa\ order \(N\). Whenever radial derivatives of the corrected quantities need to be computed (for instance, \(\dd p_{\varphi}/\dd r\) in Eq.~(\ref{eq:pa_pr})), this is performed numerically on the grid. Finally, the time \(t\) and the orbital phase \(\varphi\) are obtained through numerical integration
\begin{subequations}
\begin{align}
t(r) &= \!\int_{r_{\mathrm{min}}}^{r_{\mathrm{max}}} \dd r \left(\!\frac{\partial H}{\partial p_{r_{*}}}\!\right)^{\!-1}, \\
\varphi(r) &= \!\int_{r_{\mathrm{min}}}^{r_{\mathrm{max}}} \dd r \left(\!\frac{\partial H}{\partial p_{\varphi}}\!\right)\!\left(\!\frac{\partial H}{\partial p_{r_{*}}}\!\right)^{\!-1}.
\end{align}
\end{subequations}

The waveform is built from the \pa\ dynamics using the same prescription as the standard \eob\ waveform model. The waveform strain \(h(t) = h_{+}(t) - i h_{\times}(t)\) can be decomposed into multipoles according to
\begin{align}\label{eq:waveform}
h(t) = \frac{1}{d_{L}} \sum_{\ell = 2}^{\ell_{\mathrm{max}}} \sum_{m=-\ell}^{\ell} h_{\ell m}(t) \, _{-2} Y_{\ell m} (\theta, \phi),
\end{align}
where \(d_{L}\) is the distance from the detector to the source, and \(_{-2} Y_{\ell m} (\theta, \phi)\) are the spin-weighted spherical harmonics for \(s = -2\). \(\ell_{\mathrm{max}}\) is the highest-order multipole which is calculated. More detailed accounts of the procedure for generating \eob\ waveforms can be found in Ref.~\citep{Damour:2014sva, Nagar:2015xqa, Pan:2013rra}. A robust implementation of the \pa\ dynamics for an arbitrary spin-aligned \eob\ Hamiltonian is presented in the Sec.~\ref{sec:implementation}.

\section{Implementation in LIGO Algorithm Library}
\label{sec:implementation}
\noindent

We have implemented the post-adiabatic (\pa) inspiral dynamics model described in Sec.~\ref{sec:theory} in \lalsuite~\citep{lalsuite} and it is available through the \hmpa\ waveform model approximant.

When this model is used, the dynamics of the binary system, starting
from the initial separation \(r_{0} = r_{\mathrm{max}}\) until some
final separation \(r_{\mathrm{min}}\) is approximated with the \pa\
procedure described in Sec.~\ref{sec:theory}. The radius at which
  the \pa\ procedure is terminated, \(r_{\mathrm{min}}\), as well as
  the size of the grid \(\dd r\), are empirically chosen to ensure
  that the faithfulness of the waveform is maximised while keeping the
  computational cost minimal. Around \(10^{4}\) waveforms were
  generated, covering the space of binary parameters and exploring the
  effects of varying these two parameters. In each case we compute 
  the unfaithfulness and choose values for these parameters that 
  ensure the fastest waveform generation while still being
  sufficiently accurate. We find that the following prescription
  satisfies these requirements:
\begin{align} r_{\mathrm{min}} =
    1.6 \; r_{\mathrm{\textsc{isco}}} \quad \text{and} \quad \dd r =
    0.3.
\end{align}%
Furthermore, the \pa\ order is a free parameter of our model, with the default
  being $8^{\rm th}$ order. Our studies show that lower orders cannot always
  achieve the desired accuracy, while higher orders incur
  computational cost without further improving the solution, or can be
  prone to numerical noise (e.g. above \(12^{\rm th}\) order).

The \pa\ approach is independent of any particular form of the
Hamiltonian, and here we focus on the \hmm\
Hamiltonian~\citep{Barausse:2009xi,Boh__2017}~\footnote{We note that in Ref.~\citep{Rettegno:2019tzh} 
the authors derived the equations of motion of the (uncalibrated) Hamiltonian used in the 
\hmm\ model. However, they employed a form of the Hamiltonian that differs 
from the one used in the \hmm\ waveform model~\citep{Boh__2017,Cotesta:2018fcv}. Thus, we could not 
take advantage of their findings.}. Our procedure is
  set up to use either analytical or numerical derivatives of the
  \eob\ Hamiltonian (e.g. \(\partial H / \partial r\)), this giving
  additional flexibility to the user. The numerical derivatives in
  Eqs.~(\ref{eq:j0}), (\ref{eq:pa_pr}), and (\ref{eq:pa_pphi}) are
  computed using an $8^{\rm th}$-order finite difference method
  \citep{Burden1989, abramowitzandstegun, Fornberg1988}, while
  numerical integration is performed using a standard cube-spline
  quadrature algorithm \citep{holmes2014, Press2007}.

In order to provide an implementation of the waveform model that is maximally efficient while preserving the faithfulness at each point in parameter space, a number of further changes are introduced to the algorithms for calculating the binary dynamics and for computing the waveform modes. These changes are summarized below:
\begin{enumerate}
\item Analytic derivatives are used both during the \pa\ routine and the final \ode\ integration (where the approximant \hmopt\ is used \citep{Boh__2017, Devine_2016,Knowles:2018hqq}). Analytic derivatives are computationally more efficient than finite-difference methods, and therefore provide valuable speedup for computing the binary dynamics.
\item A larger integration step is used for the final part of the dynamics calculation (using the \hmopt\ model), which speeds up the \ode\ integration significantly.
\item Following Refs.~\citep{Boh__2017, Devine_2016}, quantities which do not vary with the mode numbers \((\ell, m)\) are pre-computed instead of being repeatedly generated during each iteration. This helps to remove a large portion of the computational overhead in building the waveform modes.
\item Finally, the waveform is computed over a non-uniformly--spaced time grid, which is comprised of the sparse grid for \pa\ approximation, and the denser grid for the final part of the dynamics (plunge and merger). This speeds up the waveform generation  considerably as waveform generation on an equally spaced grid is expensive. To obtain the final modes on an equally spaced grid we follow the interpolation approach described in~\citep{Cotesta:2020qhw}.

\end{enumerate}
\begin{figure}[t]
	\includegraphics*[scale=0.94]{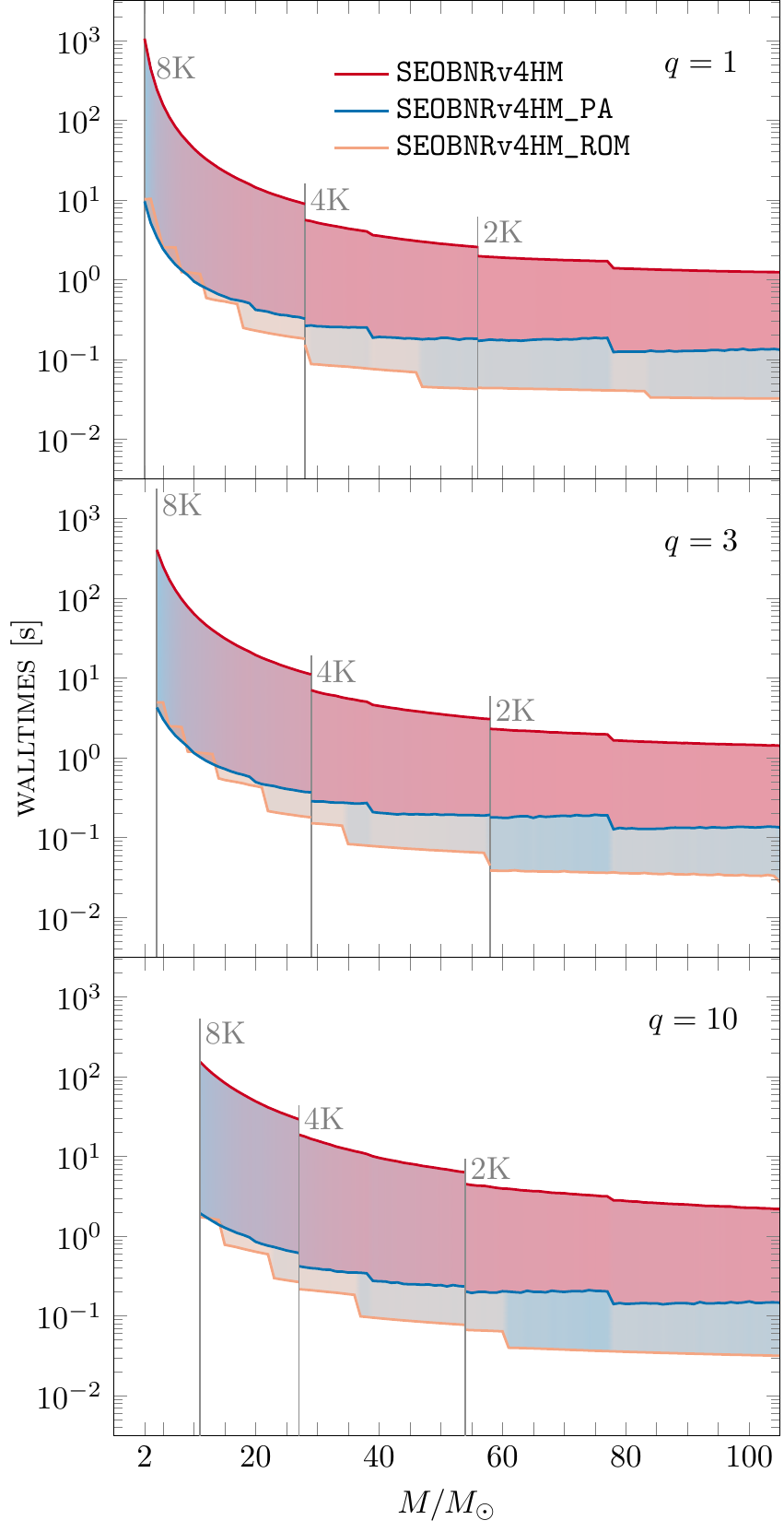}
	\caption{Benchmark of the \hmpa\ waveform model against 2 other well known and commonly used models. Compared to the \hmm\ model, the post-adiabatic model is between \(10\) and \(10^{2}\) times faster depending on the total mass (for a starting frequency of \SI{10}{\hertz}). The frequency-domain \hmrom\ model is faster for high total mass \(M\), but the two models have near-equal performance in the low-total-mass regime. In these tests, the compact objects have spins \(\chi_{1} = 0.8\) and \(\chi_{2} = 0.3\), and the sampling rate has been chosen so that it is large enough to resolve the \((5, 5)\) mode for large total mass, but also to never exceed \SI{8192}{\hertz}.}\label{fig:walltimes_plot}
\end{figure}

\begin{figure}[t]
	\includegraphics*[scale=1]{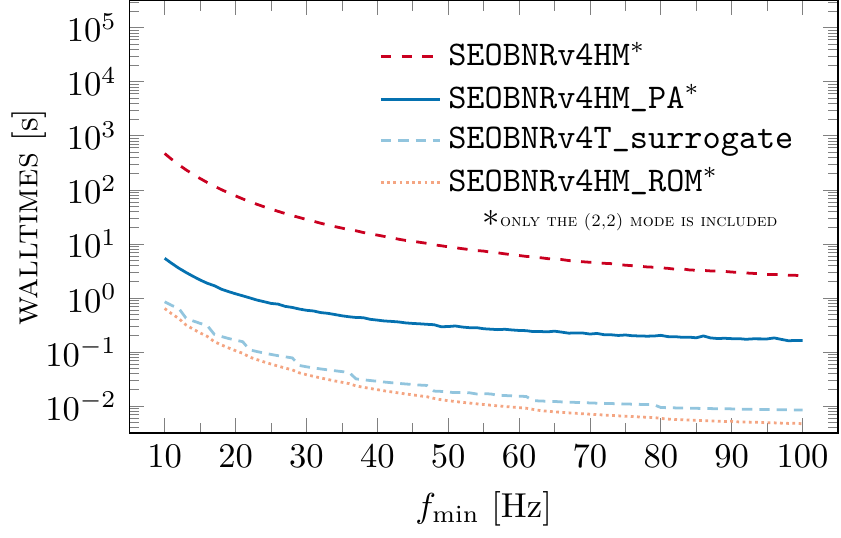}
	\caption{Benchmark of the \hmpa\ waveform model against \hmm\ and the surrogate model \sur\ for a \(1.4 M_{\odot} + 1.4 M_{\odot}\) binary with no spins for a range of starting frequencies \(f_{\mathrm{min}}\). Since the \sur\ model only includes the \((\ell, |m|) = (2, 2)\) mode, the other models were modified to only compute this mode. All waveforms were sampled at \SI{8192}{\hertz}.}\label{fig:sur_timings}
\end{figure}

\begin{figure}[t]
	\includegraphics*[scale=0.94]{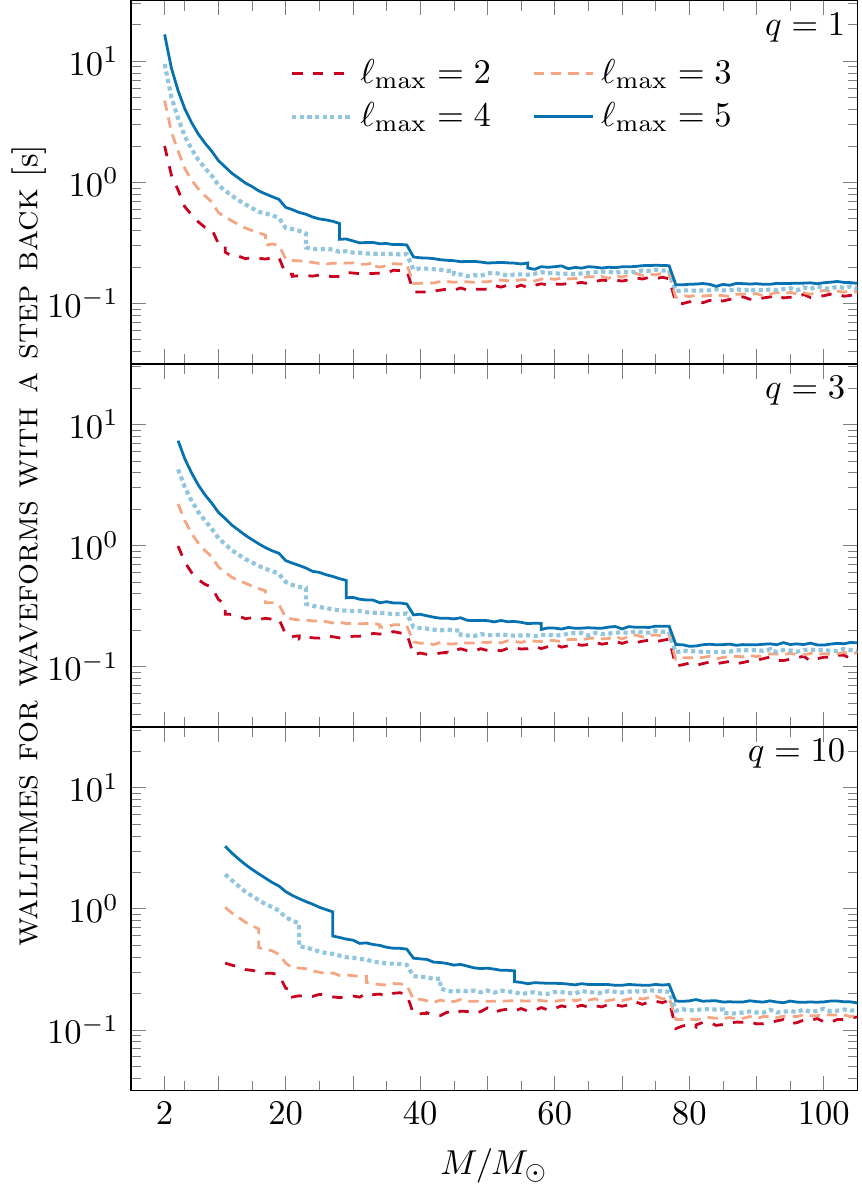}
	\caption{Benchmark of the \hmpa\ waveform with different highest-order modes allowed. For each mass ratio \(q\) and \((\ell, m)\) mode, the initial frequency was modified so that this mode is included at \(f = \SI{20}{\hertz}\). The compact objects have spins \(\chi_{1} = 0.8\) and \(\chi_{2} = 0.3\), and the sampling rate has been fixed so that it is large enough to resolve the (5, 5) mode for large total mass, but also to never exceed \SI{8192}{\hertz}.}\label{fig:modes_timings}
\end{figure}

\begin{figure*}[t]
	\includegraphics*[scale=1]{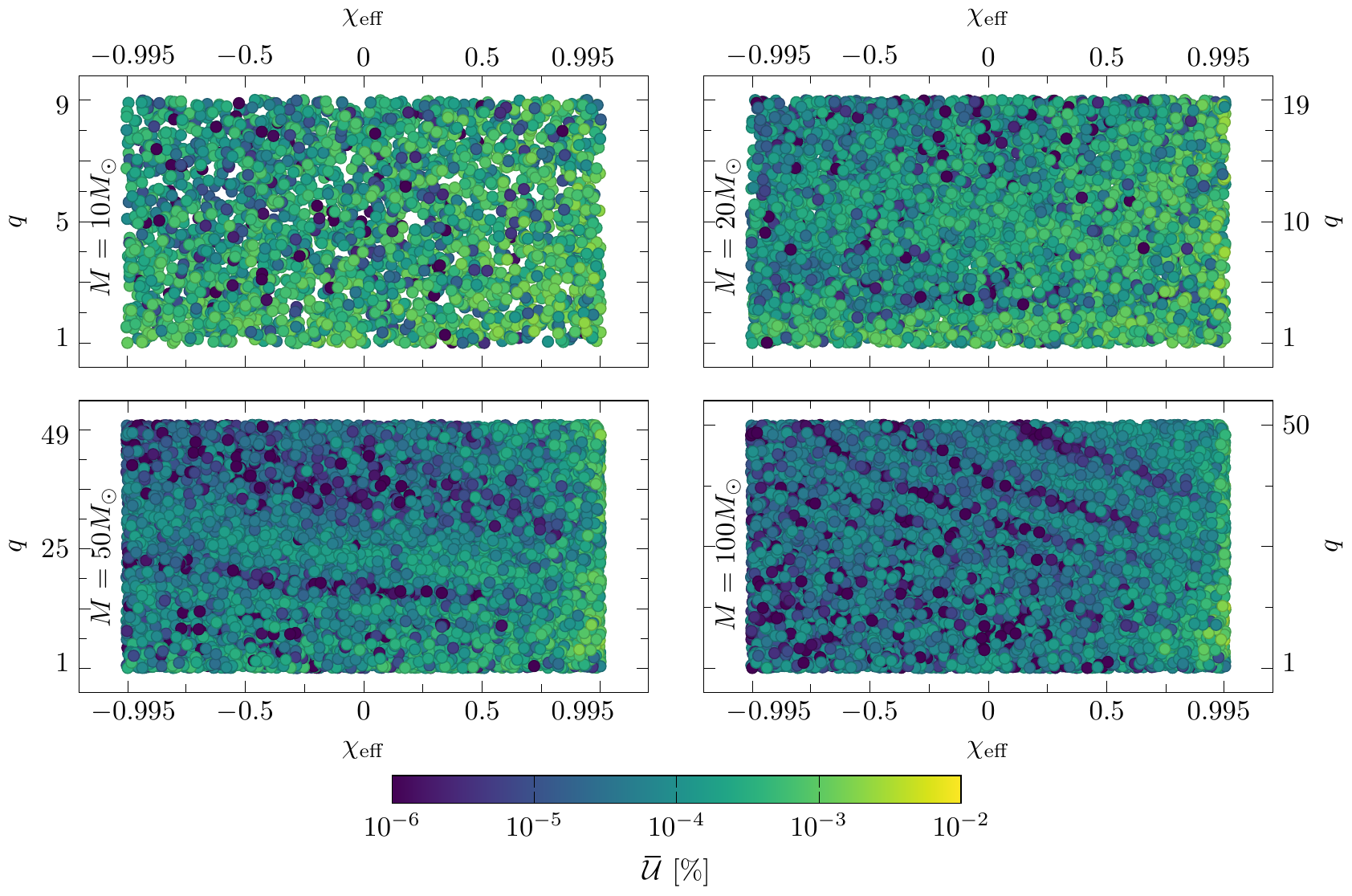}
	\caption{Unfaithfulness \(\overbar{\mathcal{U}}\) (in percent) between the \hmpa\ and \hmm\ models as a function of mass ratio \(q\) and effective spin \(\chieff\) for 4 different values of the total mass \(M\).}\label{fig:faithfulnessM}
\end{figure*}

\subsection{Computational performance of the model}
\noindent
The \hmpa\ model was benchmarked against other relevant waveform models which are available in \lalsim: against \hmm\ \citep{Cotesta:2018fcv}, \hmrom\ \citep{Cotesta:2020qhw}\, and \sur\ \citep{Lackey:2018zvw} in computational efficiency, and against \hmm\ in accuracy. Figure~\ref{fig:walltimes_plot} shows the time for generating a waveform for total masses between 2 and \(100 M_{\odot}\), with starting frequency of \SI{10}{\hertz}, and for three values of the mass ratio, \(q = \{1, 3, 10\}\). The \hmpa\ model performs significantly faster than the \hmm\ model across all values of the total mass \(M\). The speedup is most significant for lower total mass (\(\sim 60 \times\)), and drops to its lowest for high total mass (\(\sim 10 \times\)). Most importantly, the speedup is substantial (\(\sim 30 \times\)) for \(M \sim 40 - 60 M_{\odot}\), where many of the events of current interest lie. For \(M \gtrsim 10 M_{\odot}\), the time to generate a waveform using \hmpa\ is less than 1 s.

Comparing to the frequency-domain \hmrom\ model, we see that, as expected, the reduced order model is faster in almost all cases, except for very low total mass (\(M \lesssim 10 M_{\odot}\)) where the two models are comparable in terms of computational cost (see Fig.~\ref{fig:walltimes_plot}).

As a further test, the \hmpa\ model was benchmarked against the \sur\ model, which only includes the \((\ell, |m|) = (2, 2)\) mode. Figure~\ref{fig:sur_timings} compares the time for generation of a waveform with the \hmpa\ and \sur\ models where only the quadrupole mode is computed, at different sampling rates, and with different starting frequencies, for a \(1.4 M_{\odot} + 1.4 M_{\odot}\) \bns\ with no spins.

Finally, we benchmark the time necessary to generate each higher-oder multipole of the model. Figure~\ref{fig:modes_timings} shows the times to generate waveforms with the \hmpa\ model for which all \((\ell, |m|)\) modes with \(\ell \leq \ell_{\mathrm{max}}\) (see the legend) are resolved at initial frequency of \(f_{0} = \SI{20}{\hertz}\). As expected, adding higher multipoles increases the Nyquist frequency and the cost for generating the waveform. The figure demonstrates the difference in time that it takes to compute each of the modes of the model. Another test demonstrated that generating additional, higher-order modes at a fixed sampling rate incurs negligible computational cost to the waveform generation.

\subsection{Accuracy of the model}
\noindent
To assess the accuracy of our model with respect to \hmm\ we use the notion of unfaithfulness as outlined below.

In general, the \gw\ signal from a non-precessing, quasi-circular \bbh\ is characterized by a total of 11 parameters: the binary companion masses \(m_{1}\) and \(m_{2}\), the (dimensionless) component spins \(\chi_{1}\) and \(\chi_{2}\), which can be aligned or anti-aligned with the orbital angular momentum, 
the orientation of the binary in the source frame \((\iota, \varphi_{0})\), the sky location of the source in the detector frame \((\theta, \phi)\), and the luminosity distance \(d_{L}\), the time of arrival \(t_{c}\), and the polarization angle \(\psi\). The detector response can be written as
\begin{align}
h(t) &= F_{+}(\theta, \phi, \psi) \, h_{+}(\bm{\lambda}, t_{c}; t) + F_{\times}(\theta, \phi, \psi) \, h_{\times}(\bm{\lambda}, t_{c}; t),
\end{align}
where \(\bm{\lambda}=(m_{1}, m_{2}, \chi_{1}, \chi_{2}, \iota, \varphi_{0}, d_{L})\) are the binary parameters and the functions \(F_{\{+, \times\}}(\theta, \phi, \psi)\) are the antenna patterns (see \citep{Sathyaprakash:1991mt, Finn:1992xs}). This can be cast as
\begin{align}
h(t) &= \mathcal{A}(\theta, \phi) \big(\cos(\kappa) \, h_{+}(\bm{\lambda}, t_{c}; t) + \sin(\kappa) \, h_{\times}(\bm{\lambda}, t_{c}; t)\big).
\end{align}
Here \(\kappa \equiv \kappa(\theta, \phi, \psi)\) is the effective polarization \citep{Capano:2013raa}, defined as
\begin{align}
\exp(i \kappa(\theta, \phi, \psi)) = \frac{F_{+}(\theta, \phi, \psi) + i F_{\times}(\theta, \phi, \psi)}{\mathcal{A}(\theta, \phi)},
\end{align}
where the overall amplitude function \(\mathcal{A}(\theta, \phi)\) is
\begin{align}
\mathcal{A}(\theta, \phi) = \sqrt{F_{+}^{2} (\theta, \phi, \psi) + F_{\times}^{2} (\theta, \phi, \psi)} \; .
\end{align}

We can now define the match between a \gw\ signal \(h_{\sig} (t)\) (which for us is \hmm) and a template waveform \(h_{\temp} (t)\) (which is \hmpa) (see \citep{Harry:2016ijz}):
\begin{align}\label{eq:faithfulnessDef}
\mathcal{F}(\iota_{\sig}, {\varphi_{0}}_{\sig}, \kappa_{\sig}) = \max_{t_{c}, {\varphi_{0}}_{\temp}, \kappa_{\temp}} \left[ \left. \frac{(h_{\sig}, h_{\temp})}{\sqrt{(h_{\sig}, h_{\sig}) (h_{\temp}, h_{\temp})}} \right|_{\substack{{m_{1,2}}_{\sig} = {m_{1,2}}_{\temp} \\ {\chi_{1,2}}_{\sig} = {\chi_{1,2}}_{\temp} \\ {\iota_{\sig} = \iota_{\temp}}}} \right],
\end{align}
where the parameters denoted with \textsc{s} (\textsc{t}) refer to the signal (template) waveform. For the purposes of this calculation, we marginalize over the phase \({\phi_{0}}_{\temp}\), the effective polarization \(\kappa_{\temp}\), and the time of arrival \(t_{c}\), and use the familiar definition for the inner product between two waveforms \citep{Sathyaprakash:1991mt, Finn:1992xs}:
\begin{align}
(x, y) \equiv 4 \; \Re \left[ \int_{f_{\mathrm{low}}}^{f_{\mathrm{high}}} \dd f \; \frac{\widetilde{x}(f) \; \widetilde{y}^{*}(f)}{S_{\mathrm{n}} (f)} \right],
\end{align}
where a \(\sim\) denotes a Fourier transform, a \(*\) denotes a complex conjugate, and finally \(S_{\mathrm{n}} (f)\) is the one-sided power spectral density (\psd) of the detector noise.

Here, we use \(f_{\mathrm{low}} = \SI{20}{\hertz}\), \(f_{\mathrm{high}} = \SI{2048}{\hertz}\), and the Advanced \ligo\ "zero-tuned high-power" design sensitivity curve \citep{Barsotti:2018}. The definition of the faithfulness in Eq.~(\ref{eq:faithfulnessDef}) depends on the signal parameters (\(\iota_{\sig}, {\varphi_{0}}_{\sig}, \kappa_{\sig}\)) and therefore allows us to work with either the maximum or the averaged \emph{unfaithfulness} (or \emph{mismatch}) \(1 - \mathcal{F}(\iota_{\sig}, {\varphi_{0}}_{\sig}, \kappa_{\sig})\):
\begin{widetext}
\begin{subequations}
\begin{align}
\mathcal{U}_{\mathrm{max}} &\equiv \max_{\{\iota_{\sig}, {\varphi_{0}}_{\sig}, \kappa_{\sig}\}}\left[1 - \mathcal{F}\right] \equiv 1 - \min_{\{\iota_{\sig}, {\varphi_{0}}_{\sig}, \kappa_{\sig}\}} \left[\mathcal{F}\right], \\
\overbar{\mathcal{U}} &\equiv \langle 1 - \mathcal{F} \rangle_{\{\iota_{\sig}, {\varphi_{0}}_{\sig}, \kappa_{\sig}\}} \equiv 1 - \frac{1}{8 \pi^{2}} \int_{0}^{2\pi} \dd\kappa_{\sig} \int_{0}^{\pi} \dd\iota_{\sig} \sin(\iota_{\sig}) \int_{0}^{2\pi} \dd {\varphi_{0}}_{\sig} \mathcal{F}.
\end{align}
\end{subequations}
\end{widetext}

\begin{figure}[b]
	\includegraphics*[scale=1]{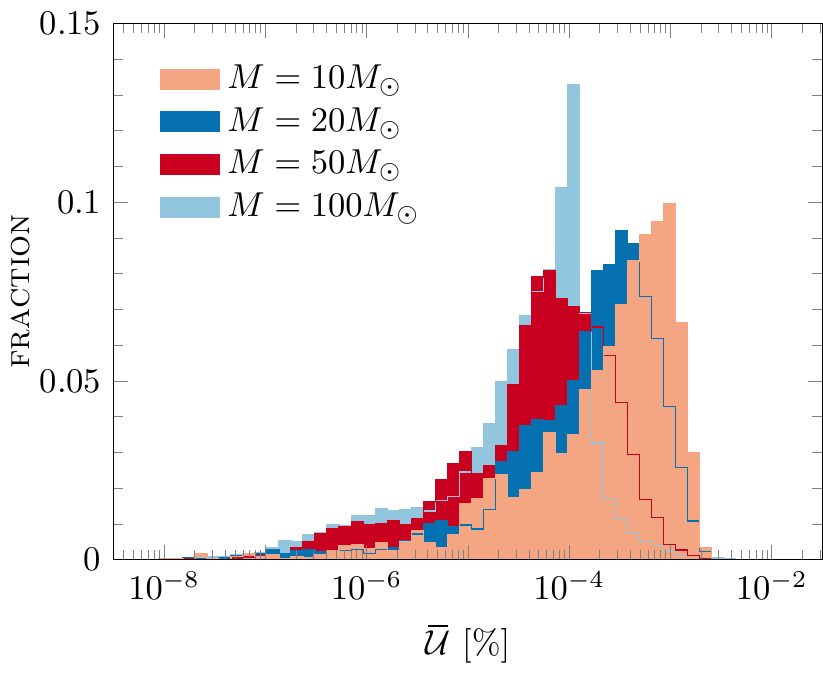}
	\caption{Histogram of the unfaithfulness \(\,\overbar{\mathcal{U}}\) (in percent) between the \hmpa\ and \hmm\ models for 4 different values of the total mass \(M\).}\label{fig:histogram}
\end{figure}

Using these definitions of the unfaithfulness, we examine the accuracy of the post-adiabatic (\pa) model in comparison with \hmm. Figure~\ref{fig:faithfulnessM} shows the maximum unfaithfulness \(\mathcal{U}_{\mathrm{max}}\) across the parameter space of the effective spin \(\chieff\) and the mass ratio \(q\) for 4 separate values of the total mass \(M\). We find that in all cases that were presented, the unfaithfulness is below \(\bigo(0.01 \%)\), which makes our model comparable in accuracy to \hmrom~\citep{Cotesta:2020qhw}. The variations of the unfaithfulness for a fixed total mass can be explained in terms of the length of the waveform and the heuristic condition which we use to transition from \pa\ inspiral to merger and ringdown.

The variations across the separate panels of Fig.~\ref{fig:faithfulnessM} can be better expressed in the form of a histogram of the average unfaithfulness \(\overbar{\mathcal{U}}\), which is shown in Fig.~\ref{fig:histogram}. The distribution for each value of the total mass \(M\) peaks at a higher value of the unfaithfulness, which is consistent with the fact that lower-mass binaries undergo a longer coalescence inside the frequency band relevant to \ligo. In all cases, however, the average unfaithfulness stays below \(\bigo(0.01 \%)\).

\begin{figure*}
	\subfloat[Marginalized 2D posterior for the component source-frame masses \(m_{1}\) and \(m_{2}\).\label{fig:inj_m1_m2}]{\includegraphics[scale=1]{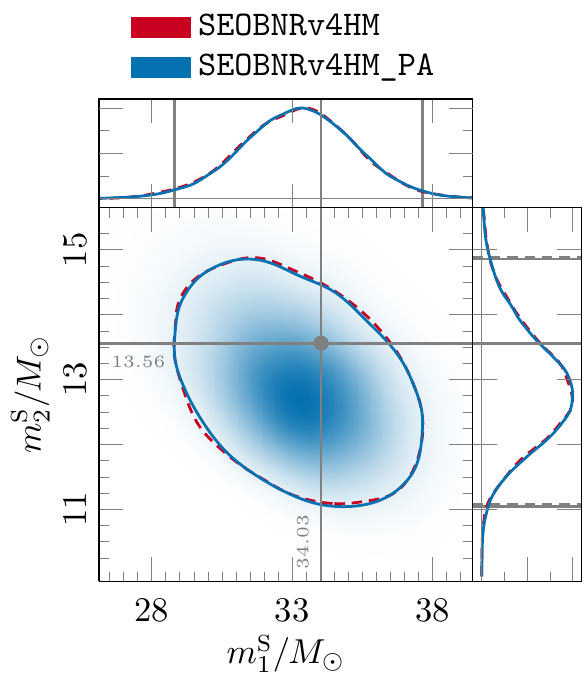}}\hfill%
	\subfloat[Marginalized 2D posterior for the mass ratio \(q\) and the effective spin \(\chieff\).\label{fig:inj_chieff_q}]{\includegraphics[scale=1]{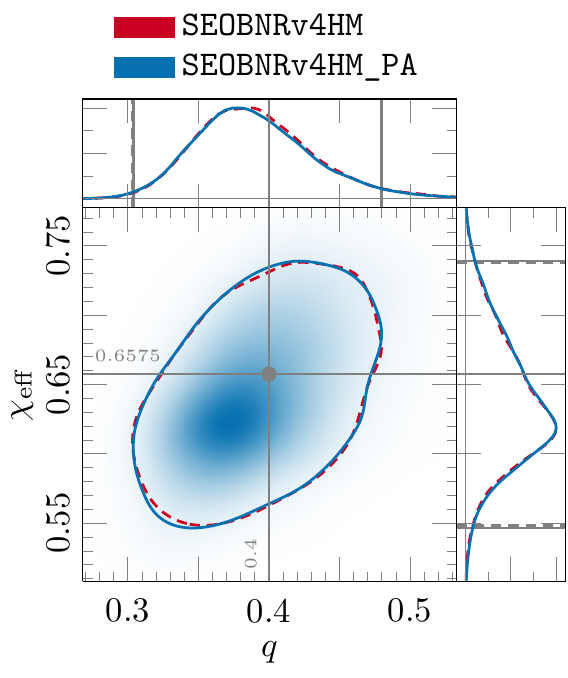}}\hfill%
	\subfloat[Marginalized 2D posterior for the luminosity distance \(d_{L}\) and the inclination of the binary \(\theta\).\label{fig:inj_dl_i}]{\includegraphics[scale=1]{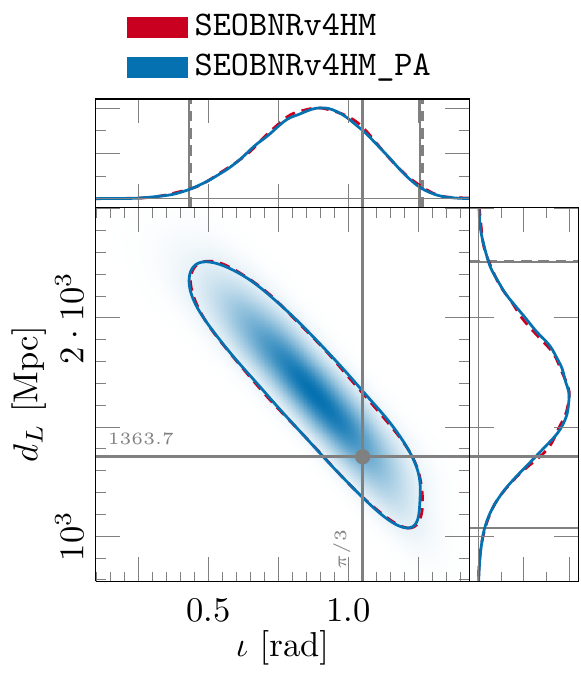}}
    \caption{Parameter estimation results for the synthetic injection. The 2-dimensional posterior plots show the 90\% confidence regions for the parameters. The grey vertical lines in the 1-dimensional plots show the projections of these confidence regions. The grey labelled lines denote the injection values. Note the excellent agreement between \hmpa\ and \hmm\ waveform models.}\label{fig:injection}
\end{figure*}

\section{Parameter estimation study}
\label{sec:pestudy}

While the previous section deals with the absolute performance of the \hmpa\ model, it is important to demonstrate that these benchmarks translate into a faster and reliable parameter estimation (\pe) analyses. For this purpose, two \pe\ studies were performed with the \hmpa\ model, as well as some of the established waveform models (\hmm\ or \hmrom). The first one is an injection study involving a synthetic signal. The data is then analysed using \hmpa\ and \hmm\ models. The second \pe\ study was performed on data from the event \event. It was analysed using the \hmpa\ and the \hmrom\ models.

In both cases, the parameter estimation was done using the Markov-chain Monte Carlo code \lalmcmc\ \citep{lalsuite}. The results of these studies are presented below in the following subsections. We also note that \pe\ studies using the \texttt{TEOBResumS} waveform  model with the \pa\ approximation were done in Refs~\citep{Akcay:2018yyh, Gamba:2020wgg, Akcay:2020qrj, Gamba:2020ljo, Breschi:2021wzr}.

\subsection{Injection-based study}
\label{sec:injection}
\noindent
The choice of parameters for this \pe\ study was made so that it could emphasise the speedup which the \hmpa\ model offers, but also to manifest higher-order modes in the signal. The injected signal has total mass \(M = 60 M_{\odot}\) with mass ratio \(1/q \approx 0.4\), aligned spins \((\chi_1, \chi_2) = (0.8, 0.3)\), at a distance of approximately 1364 Mpc (corresponding to signal-to-noise ratio (\snr) of \(\sim 22\)) and at an inclination of \(\pi/3\). The injection was made using the \hmm\ model, while the subsequent analysis was performed using both the \hmm\ and the \hmpa\ models in order to judge the performance of the post-adiabatic (\pa) model.

The results of the parameter estimation analysis can be summarized in the series of marginalized 2-dimensional posterior plots in Fig.~\ref{fig:injection}. Figure~\ref{fig:inj_m1_m2} shows the plot for the component source-frame masses \(m_{1}^{\textsc s}\) and \(m_{2}^{\textsc{s}}\); Fig.~\ref{fig:inj_chieff_q} shows the plot for the mass ratio \(q\) and the effective spin \(\chieff\). Finally, Fig.~\ref{fig:inj_dl_i} shows the posterior for the luminosity distance \(d_{L}\) (cf. Eq.~(\ref{eq:waveform})) and the inclination angle \(\iota\). The plots demonstrate that the distributions of the posterior samples are in very good agreement between the two models, which demonstrates the reliability of \hmpa\ in \pe\ studies. The Jensen-Shannon (\js) divergence between the 1-dimensional posteriors which are shown in Fig.~\ref{fig:injection} is always at \(\bigo\!\left(10^{-3}\right)\) or below, which is consistent with the effects of stochastic sampling on the recovered quantities \citep{Romero-Shaw:2020owr}.

\begin{figure*}
	\subfloat[Marginalized 2D posterior for the component source-frame masses \(m_{1}\) and \(m_{2}\).\label{fig:event_masses}]{\includegraphics[scale=1]{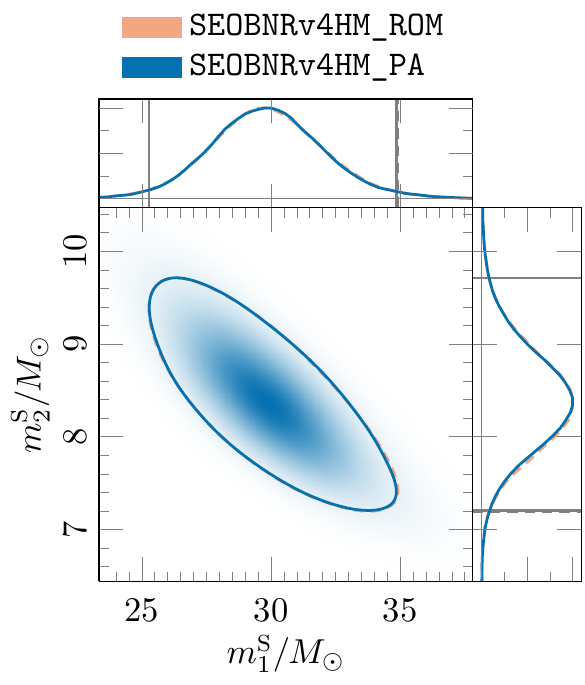}}\hfill%
	\subfloat[Marginalized 2D posterior for the mass ratio \(q\) and the effective spin \(\chieff\).\label{fig:event_qchi}]{\includegraphics[scale=1]{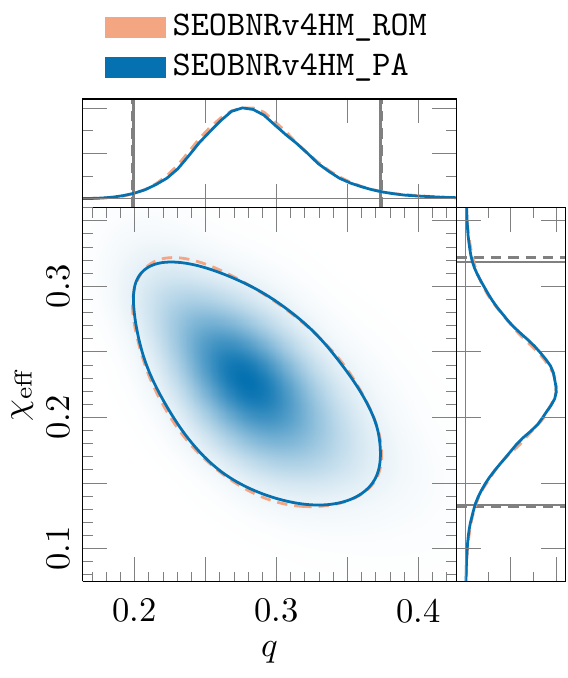}}\hfill%
	\subfloat[Marginalized 2D posterior for the luminosity distance \(d_{L}\) and the inclination of the binary \(\iota\).\label{fig:event_jth}]{\includegraphics[scale=1]{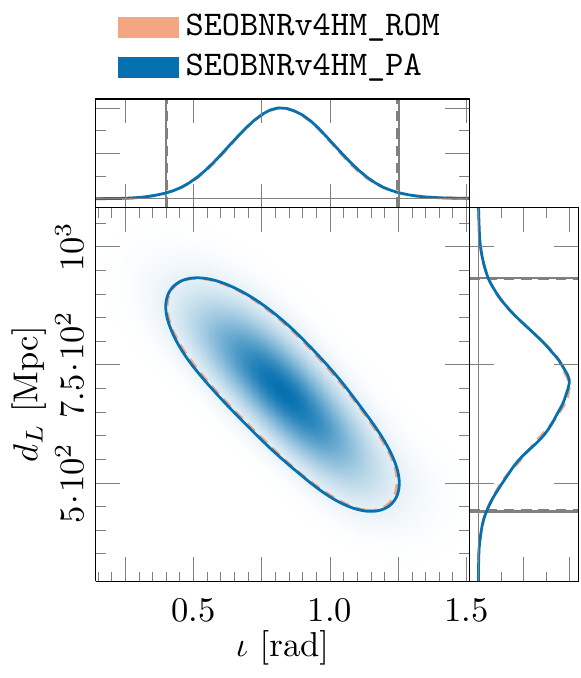}}
    \caption{Parameter estimation results for the event \event. The 2-dimensional posterior plots show the 90\% confidence regions for the parameters. The grey vertical lines in the 1-dimensional plots show the projections of these confidence regions. There is an excellent agreement between the results with the \hmpa\ and the \hmrom\ models.}\label{fig:event}
\end{figure*}

\subsection{GW190412 parameter estimation}
\label{sec:event}
\noindent
It is more interesting to demonstrate the viability of the \hmpa\ model on a real \gw-event. The \event\ from the LIGO O3a catalogue is a good candidate for such a study due to its parameters -- total mass of \(\lesssim 40 M_{\odot}\) (\(m_1 \approx 30 M_{\odot}\) and \(m_2 \approx 8 M_{\odot}\), asymmetric mass ratio of about 3, and dimensionless spin of the massive companion between 0.22 and 0.6 \citep{LIGOScientific:2020stg, Zevin:2020gxf, Islam:2020reh, Colleoni:2020tgc}. The total mass and mass ratio suggest that waveforms for this event would be computationally much more efficient using the \hmpa\ model -- which means it is particularly suitable to study this event. The analyses were performed with both the \hmpa\ and \hmrom\ models, since any analysis with the \hmm\ model would have been impractical in terms of the time required to complete it.

The 2-dimensional marginalized posterior plots are shown in Fig.~\ref{fig:event}. Figure~\ref{fig:event_masses} shows the posterior for the component source-frame masses \(m_{1}^{\textsc s}\) and \(m_{2}^{\textsc{s}}\) of the binary; Fig.~\ref{fig:event_qchi} shows the plot for the mass ratio \(q\) and the effective spin \(\chieff\); and finally, Fig.~\ref{fig:event_jth} shows the plot for the luminosity distance \(d_{L}\) and the inclination angle \(\iota\). The plots demonstrate the incredibly good agreement between the \hmrom\ and the \hmpa\ models, which is further confirmed by the \js\ divergence between the samples, which is below the \(\bigo\!\left(10^{-3}\right)\) level.

In conclusion for this section, the \hmpa\ model may be reliably used for parameter estimation analyses in place of models like \hmm, with no evident caveats which could hinder the results of such analyses. Furthermore, the average speedup in the generation of samples for the \textsc{mcmc} chains was found to be around 1 order of magnitude. The reason that this is less compared to the speedup demonstrated in Fig.~\ref{fig:walltimes_plot} is the fact that for the \pe\ studies, the waveforms are generated using starting frequency \(f_{0} = \SI{20}{\hertz}\).

\section{Tests of General Relativity}
\label{sec:tgr}

In previous sections we have seen that \hmpa\ is a robust, accurate and fast alternative to \hmm\ and can therefore be used as a drop-in replacement. An interesting further application of this is to use the \hmpa\ model for tests of General Relativity, where the \hmm\ model was previously used as a baseline \gr\ model. In particular, we consider a parametrized black hole (\bh) ringdown test that measures the deviations of quasi-normal mode emission from predictions of \gr. We summarize the test briefly here, see~\citep{Abbott:2020jks,Ghosh:2021mrv} for more details.
In \gr, the no-hair conjecture predicts that the physical properties of a (uncharged) \bh\ are completely determined by its mass and spin. Consequently, the quasi-normal modes (\qnm) that describe the gravitational waves emitted by a perturbed \bh\ are also uniquely determined by its mass and spin. Thus one can check the validity of \gr\ by measuring or constraining any deviations in the complex \qnm\ frequencies. 
The \pSEOB\ ringdown analysis uses a parameterised version of a full inspiral-merger-ringdown \gw\ signal model to measure and constrain the (complex) \qnm\ frequencies. Consequently, unlike other ringdown studies restricted to the post-merger signal~\citep{Isi:2019aib,Giesler:2019uxc,Carullo:2018gah,Carullo:2019flw}, the \pSEOB\ analysis makes use of the entire signal power and does not suffer from the ambiguity of a ringdown start-time definition. In \pSEOB, one starts with the \hmm\ model, but then introduces deviations \(\delta f_{lm0}\) and \(\delta \tau_{lm0}\) (which are treated as free parameters) to the \qnm\ frequency and damping time, so that
\begin{align}
f_{lm0} &= f_{lm0}^{\rm \textsc{gr}} \left(1+\delta f_{lm0}\right), \\
\tau_{lm0} &= \tau_{lm0}^{\rm \textsc{gr}} \left(1+\delta \tau_{lm0}\right),
\end{align}
and the ringdown signal is different from the \gr\ prediction. Here, \(f_{lm0}^{\rm \textsc{gr}}\) and \(\tau_{lm0}^{\rm \textsc{gr}}\) are computed from final mass and spin as predicted by \nr\ fitting formulae. The goal of the test is to infer the values of the \((\delta f_{lm0}$, $\delta \tau_{lm0})\) by doing full parameter estimation.

\begin{figure}[t]
    \centering
    \includegraphics[scale=1]{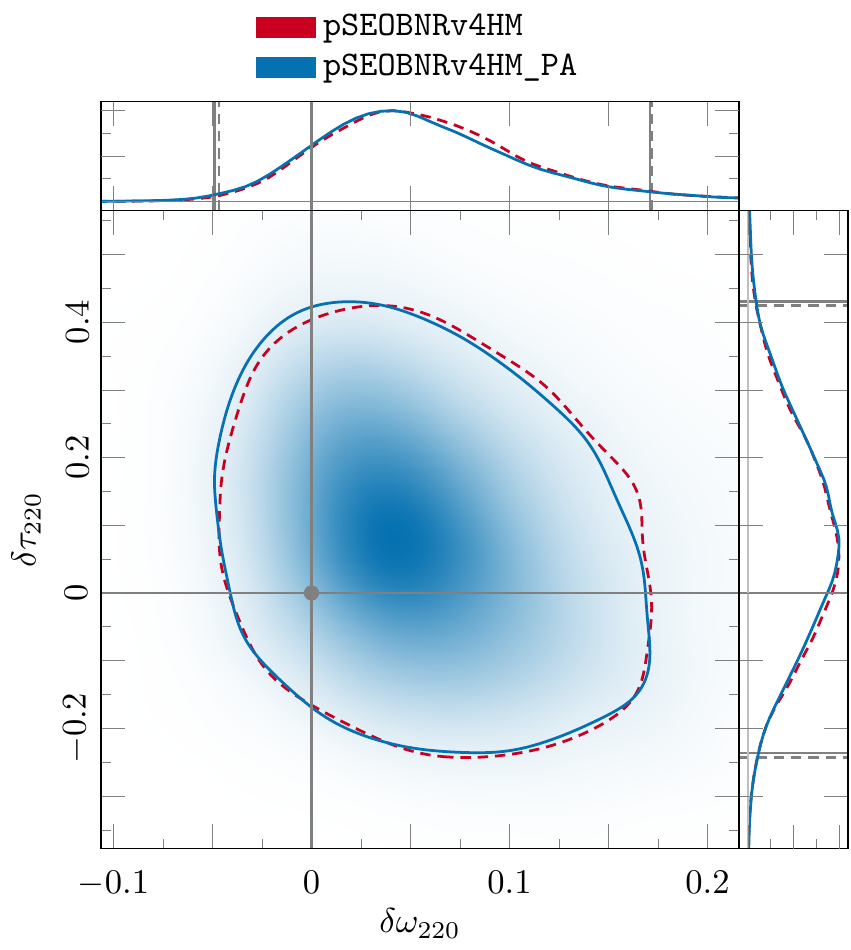}
    \cprotect\caption{Posterior distributions for the fractional deviations \(\delta f_{220}\) and \(\delta \tau_{220}\) in the damping time and frequency as recovered by using \hmm\ and \hmpa. The 2-dimensional posterior plot shows the 90\% confidence regions for the parameters. The grey vertical lines in the 1-dimensional plots show the projections of these confidence regions. The thick grey lines denote the values predicted by \gr.}
    \label{fig:tgr}
\end{figure}

Performing full parameter estimation taking into account deviations from \gr\ is a challenging problem, since it involves sampling a higher-dimensional space (two new parameters for every \((\ell, m)\) mode considered), which means any improvements to waveform speed can lead to even bigger impact on the overall runtime. To check the speed and accuracy of using \hmpa\ for this application, we compare it to the results obtained with \hmm\ for the first gravitational wave event ever detected, \tgrevent. We choose this event as it has significant signal-to-noise ratio (\snr) in the merger-ringdown regime. For a high (quasi) equal-mass binary like \tgrevent, contributions from higher multipoles of the \gw\ signal are expected to be negligible. Hence, we restrict our analysis to just the least-damped dominant \qnm, i.e., \((f_{220}, \tau_{220})\), keeping the other \qnm s fixed at their nominal \gr\ values. In Fig.~\ref{fig:tgr} we show the posterior distributions for the fractional deviations in the damping time and frequency as recovered by using \hmm\ and \hmpa. It is evident that the posteriors are in extremely good agreement with the results showing consistency with \gr. More quantitatively, the Jensen–Shannon~(\js) divergence between the 1-dimensional posteriors is $\mathcal{O}(10^{-3})$ for \(\delta f_{220}\) and \(\delta \tau_{220}\), which is again within the range of what is expected due to stochastic sampling. We find a similarly good level of agreement for the other parameters. Finally, we find a speed-up of \(\sim 10\) times when using \hmpa\ instead of \hmm\ as the base \gr\ model, significantly accelerating inference. With focus shifting from analyses of individual events to population studies, demands on computational resources and person-power are ever-increasing, as demonstrated by large-scale studies in \lvk\ catalog papers \citep{LIGOScientific:2018mvr, Nitz:2019hdf}. Hence, such increases in computational efficiency is immensely important for the future of \gw\ data analysis.

\section{Discussion and conclusions}
\label{sec:conclusion}

We developed the \hmpa\ model, which combines the multipolar
  \eob\ \nr-informed model \hmm\ with the post-adiabatic (\pa)
  approach for solving the (spin-aligned) binary dynamics (developed
  and used in the \texttt{TEOBResumS} models~\citep{Damour:2012ky,Nagar:2018gnk, Rettegno:2019tzh, 
Riemenschneider:2021ppj}). The resulting model is computationally cheap (at most on the
order of seconds and less than 1 second for most of the parameter
space) and highly accurate (unfaithfulness less than \(\mathcal{O}
(10^{-3})\) when benchmarked against the \hmm\ model). Therefore, it
can be used as a drop-in replacement for parameter estimation studies
and tests of \gr\ for many \gw\ events where the use of the \hmm\
model would have been impractical.

In Sec.~\ref{sec:theory} we presented the \eob\ formalism and the derivation of the \pa\ equations, together with a discussion on the specifics of transitioning from the \pa\ regime to the merger part of the dynamics. Section~\ref{sec:implementation} describes the practical implementation of this model in the \lalsim\ library of waveform models, and provides the important benchmarks in terms of speed and accuracy of the \hmpa\ model. In particular, we demonstrated that the \hmpa\ model provides a speed-up of 10 to 100 times compared to the \hmm\ waveform model. Furthermore, we showed that the new model is accurate at a level which allows us to use it for the purposes of \ligo\ data analysis.

Section~\ref{sec:pestudy} makes use of the results of the previous section and illustrates the use of the \hmpa\ model in 2 separate parameter estimation studies: recovering an injected synthetic signal and applying the model to analyse the event \event. In both cases, we find an extremely good agreement between the results obtained with our model and other established models. Compared to \pe\ with the \hmm\ model, the study with \hmpa\ completes around 10 times faster. Finally, Sec.~\ref{sec:tgr} depicts the use of the waveform model to a test of \gr\ and discusses the importance of the model for this type of analyses. We find that there is an excellent agreement between the results obtained using the \hmpa\ and the established \hmm\ and \hmrom\ models.

There are several interesting future directions that can be pursued to apply  the \pa\ approximation to other 
\hmm\ models, notably models that include tidal deformabilities for binary neutron stars~\citep{Hinderer:2016eia,Steinhoff:2016rfi}, 
for which, currently, only the surrogate method has been applied~\citep{Lackey:2018zvw} (see Refs.~\citep{Akcay:2018yyh, Nagar:2018plt} for 
\pa\ models with tidal effects using \texttt{TEOBResumS}). A more important but challenging extension would involve 
the precessing \hmmp\ Hamiltonian~\citep{Ossokine:2020kjp}, which would allow the model to be used more efficiently and for a broader range of \gw\ events 
when doing parameter-estimation studies. Finally, the \pa\ approximation could also be extended to binary systems on eccentric dynamics. The main challenges 
with spin-precession and eccentricity is the presence of time scales beyond the orbital and radiation-reaction ones.

\section*{Acknowledgments}

The authors are grateful to Alessandro Nagar and Pierro Rettegno for helpful discussions on the topic of the post-adiabatic approximation, as well as for 
sharing their post-adiabatic code at the initial stages of this project. The data-analysis studies in this work were obtained with the HPC clusters {\tt Hypatia} and {\tt
  Minerva} at the Max Planck Institute for Gravitational Physics. 
This research has made use of data, software and/or web
tools obtained from the Gravitational Wave Open Science
Center (\href{https://www.gw-openscience.org/about/}{https://www.gw-openscience.org}), a service of \ligo\
Laboratory, the \ligo\ Scientific Collaboration and the Virgo
Collaboration. \ligo\ is funded by the U.S. National Science
Foundation (\textsc{nsf}). Virgo is funded by the French Centre National de
Recherche Scientifique (\textsc{cnrs}), the Italian Istituto Nazionale
della Fisica Nucleare (\textsc{infn}) and the Dutch Nikhef, with contributions by Polish and Hungarian institutes. 
This paper is based upon work supported by \textsc{nsf}’s \ligo\ Laboratory which is a major facility fully funded 
by the National Science Foundation.

\bibliographystyle{apsrev4-1mod.bst}
\bibliography{references}

\end{document}